\documentclass[12pt]{JHEP3}
\usepackage{amsmath,amssymb}

\def\beq{\begin{equation}}
\def\eeq{\end{equation}}
\def\bea{\begin{eqnarray}}
\def\eea{\end{eqnarray}}
\def\nn{\nonumber}

\def\d{\mathrm{d}}
\def\ap{\alpha^{\prime}}
\def\kt{\tilde{k}}

\author{ Steven Abel, Chong-Sun Chu and Mark Goodsell \\  
Centre for Particle Theory, University of Durham, Durham, DH1 3LE, UK \\
E-mail: \email{s.a.abel@durham.ac.uk}, \email{chong-sun.chu@durham.ac.uk},  \email{m.d.goodsell@durham.ac.uk} }

\title {Noncommutativity from the string perspective: 
modification of gravity at a $mm$ without $mm$ sized extra dimensions}

\abstract{We explore how the IR
pathologies of noncommutative field theory 
are resolved when the theory is realized as open strings 
in background $B$-fields: essentially, since the IR singularities 
are induced by UV/IR mixing, string theory brings them under control in much 
the same way as it does the UV singularities.
We show that at intermediate scales (where 
the Seiberg-Witten limit is a good approximation) the 
theory reproduces the noncommutative field theory with 
all the (un)usual features such as UV/IR mixing, but that 
outside this regime, in the deep infra-red, the theory 
flows continuously to the commutative theory and normal Wilsonian 
behaviour is restored. The resulting 
low energy physics resembles normal commutative physics, 
but with additional 
suppressed Lorentz violating operators. We also show that the 
phenomenon of UV/IR mixing occurs for the graviton as well, 
with the result that, in configurations where Planck's constant receives
a significant one-loop correction (for example brane-induced 
gravity), the distance scale below which gravity becomes non-Newtonian can 
be much greater than any compact dimensions. 
}

\preprint{IPPP/06/43\\DCPT/06/86\\hep-th/0606248}

\begin{document}

\maketitle

\section{Introduction}

Gauge theories in which the coordinates are noncommuting,
\begin{equation}
[x^{\mu},x^{\nu}]=i\theta^{\mu\nu}
\end{equation}
are interesting candidates for
particle physics, with curious properties (for general reviews of
noncommutative gauge theories see
refs.~\cite{Seiberg:1999vs,Douglas:2001ba,Szabo:2001kg}). 
One whose consequences we would like to understand a little better 
is ultra-violet(UV)/infra-red(IR) mixing ~\cite{Minwalla:1999px,Matusis:2000jf}. 
This is a phenomenon which gives rise to various pathologies in the 
field theory, making it, at best, difficult to understand. 
In this paper we set about examining UV/IR mixing from the point of view
of string theory with a background antisymmetric tensor ($B^{\mu\nu}$) field,
which provides a convenient UV (and hence IR) 
completion. Along the way, as well as seeing how the 
pathological behaviour is smoothed out, we will outline the
characteristic phenomenology of this general class of theories in the 
deep IR (i.e. at energy scales lower than those where noncommutative 
field theory is a good description): 
they resemble the $B=0$ theories but with Lorentz 
violating operators which 
can be taken parametrically and continuously to zero by reducing the 
VEV of $B^{\mu\nu}$. As a 
bi-product we also show that the UV/IR mixing phenomenon extends to the 
gravitational sector (although a field theoretical interpretation 
for UV/IR mixing in gravity is difficult to obtain). This allows 
the curious possibility that gravity may be non-Newtonian on much
longer length scales than those associated with the compact dimensions. 

Because UV/IR mixing, and the particular problems and phenomena 
to which it gives rise, are rather subtle, we begin now 
with a detailed discussion of exactly what questions
we would like the string theory to answer, after which we restate our findings 
in more precise terms. UV/IR mixing has its origin in the fact that the 
commutation relations intertwine   
large and small scales. At the simplest level, in a gedanken experiment 
where $x_1$ and $x_2$ do not commute, the uncertainty relation $\Delta x_1 \Delta
x_2 \sim i\theta^{12} $ together with the usual Heisenberg uncertainty 
$\Delta x_1
\Delta p_1 \sim i $ imply $\Delta x_2 \sim -\theta^{12} \Delta p_1 $: 
short distances in the $1$ direction are connected to small
momenta in the 2 direction and vice versa. 
At the field theory level, this intertwining of UV and IR 
leads to the infamous phenomenon of UV/IR mixing in the non-planar Feynman 
diagrams:
nonplanar diagrams are regulated in the UV but diverge in the IR.
Essentially, contrary to the standard picture of the Wilsonian effective action, 
heavy modes do not decouple in 
the IR so that, for example, trace U(1) factors of
the gauge group run to a free field theory in the IR even if there are
no massless excitations
~\cite{Armoni:2000xr,Khoze:2000sy,Armoni:2001br,Hollowood:2001ng,Khoze:2004zc}.

The agent responsible for these unusual and 
challenging features of noncommutative gauge field theories
is the Moyal star product,
\begin{equation}
(\phi * \varphi) (x) \equiv \phi(x)\  e^{{i\over 2}\theta^{\mu\nu}
\stackrel{\leftarrow}{\partial_\mu}
\stackrel{\rightarrow}{\partial_\nu}} \  \varphi(x),  \label{stardef}
\end{equation}
used in their definition. It induces a phase factor 
$\exp{\frac{i}{2} k.\theta.q}$ 
in the vertices, where $k$ is an external momentum and $q$ is a loop-momentum. 
This oscillating phase regulates the nonplanar diagrams in the UV, which 
can most easily be expressed using Schwinger integrals: for example the one-loop 
contribution to vacuum polarization takes the form 
(c.f. \cite{Armoni:2000xr,Khoze:2000sy,Armoni:2001br,Hollowood:2001ng,Alvarez-Gaume:2003,Alvarez-Gaume:2003kx})
\begin{equation}
\Pi_{\mu\nu}(k)\sim \int \frac{dt}{t} e^{-\frac{\tilde{k}^2}{4t}}\ldots
\end{equation}
where $\tilde{k}^\mu = \theta^{\mu\nu}k_\nu $ and the ellipsis stands
for factors independent of $\tilde{k}$. The exponential factor in the integrand
is a regulator at $t\sim \tilde{k}^2 \sim k^2 /M_{NC}^4$, where we
define the generic noncommutativity scale by $\theta^{\mu\nu} =
{\cal{O}}(M^{-2}_{NC})$. Thus the diagram, which without this factor
would be UV divergent, is regulated but only so long as $\tilde{k}\neq
0$. The result is that the UV divergences of the planar diagrams
reappear as IR poles in $\tilde{k}$ in the nonplanar diagrams. 

These divergences are problematic. First they signal a discontinuity
because the $\tilde{k}\rightarrow 0$ limit of the integrals is not
uniformly convergent: physics in the limit $\theta\rightarrow 0$ does
not tend continuously to the commutative theory.
Moreover they lead to
alarming violations of Lorentz invariance. For example, the
lightcone is generally modified to a lightwedge
\cite{Alvarez-Gaume:2003,lc1}. This is 
in sharp disagreement with observation. 
Furthermore in noncommutative gauge theory, the trace U(1) photon has
a polarization tensor  given by \cite{Matusis:2000jf}
\begin{equation}
\Pi_{\mu\nu} = \Pi_1(k^2,\tilde k^2) \, 
\left( k^2 g_{\mu\nu} - k_\mu k_\nu \right)
+ \Pi_{2} (k^2, \tilde{k}^2)\, \frac{\tilde{k}_{\mu}\tilde{k}_{\nu}}{\tilde{k}^4}
\, ,
\label{Dispersion}\end{equation}
where the additional term $\sim \Pi_{2}$ is 
multiplied by a  Lorentz violating tensor structure. It is
absent in supersymmetric theories \cite{Matusis:2000jf}, but since
supersymmetry is broken, we expect it to be at least of order
$M_{SUSY}^2$ times by some factor logarithmic in $\tilde{k}$ (where
$M_{SUSY}$ is a measure of the supersymmetry breaking). The result is a
mass of order $M_{SUSY}$ for certain polarizations of the trace-U(1)
photon while other polarizations remain massless \cite{Jaeckel:2005wt}.
Gymnastics are then required to prevent this trace U(1) photon mixing with the
physical photon.

\if
To see these properties in more detail we can introduce a universal UV cut-off 
in all Schwinger integrals. 
This is sufficient to remove the $\theta\rightarrow 0$ discontinuity
since now $\tilde{k}^2$ is accompanied by the UV cut-off itself; indeed
$\tilde{k}^2$ is everywhere replaced by $\tilde{k}^2+\Lambda_{UV}^{-2}$.
The $\Pi_2$ term for example is then of the form 
\[ \Pi_{2} (k^2, (\tilde{k}^2+\Lambda_{UV}^{-2}))
\, \frac{\tilde{k}_{\mu}\tilde{k}_{\nu}}{(\tilde{k}^2+\Lambda_{UV}^{-2})^2}. \]
It is now clear that the $\Lambda\rightarrow 0$ limit and the
$\theta\rightarrow 0$ limit do not commute.
\fi 

Clearly then, the outlook from the perspective of field theory is gloomy; 
because the IR singularities are 
a reflection of the fact that field theory is UV divergent,
any attempt to resolve them without modifying the UV
behaviour of the field theory is doomed. 
With this understanding, the general expectation is for a more encouraging 
picture in a theory with a UV completion, such as string theory. 
A more precise argument is the following.  
First it is easy to appreciate that, without an explicit 
UV completion, noncommutative field theory is unable
to describe physics in the IR limit. As noted in
ref.\cite{Abel:2006sp} and in the specific context of string theory
in ref
\cite{Gomis:2000bn,Liu:2000ad}, UV/IR mixing imposes a
IR cut-off given by 
$|k|> \Lambda_{IR} = \frac{M_{NC}^2}{\Lambda_{UV}}.$
Inside the range 
\beq \Lambda_{IR}^{ij} \sim \frac{1}{|\theta_{ij}|\Lambda_{UV}} < |k|
 < \Lambda_{UV}, \eeq
the field theory behaves in a Wilsonian manner, in the sense that modes with
masses greater than the UV cut-off do not (upto 
small corrections) affect the physics there. However outside this 
range the Wilsonian approach breaks down because modes above
$\Lambda_{UV}$ affect the physics below $\Lambda_{IR}$.
Indeed this inequality makes it impossible to make statements about 
either the $\theta^{ij}\rightarrow 0$ limit 
or the ${\tilde{k}} \rightarrow 0$ limit within field theory. 
{\it In other words, a UV completion is needed not only to describe physics above
$\Lambda_{UV}$ but also physics below $\Lambda_{IR}$}, and in
particular to discuss the existence or otherwise of discontinuities
there. The picture is most obvious in the context of running of gauge
couplings. Between $\Lambda_{IR}$ and
$\Lambda_{UV}$ the effective action accurately describes the running of
the trace U(1) gauge coupling regardless of what happens above
$\Lambda_{UV}$. Below $\Lambda_{IR}$, UV physics intervenes. For example
a period of power law running due to KK thresholds in the UV is mirrored
by the "inverse" power law running in the IR. 
Now, the precise UV completion
may take various forms, but suppose for example that it acts like 
a simple exponential cut-off, $e^{-\frac{\Lambda_{UV}^2}{4t}}$,
in the Schwinger integral. The planar diagrams are regulated in the 
usual manner, 
but the nett effect of the noncommutativity for the nonplanar diagrams 
is that the UV cut-off 
$\Lambda_{UV}^2$ is replaced by
$\Lambda_{eff}^2 =1/(\tilde{k}^2 +\Lambda_{UV}^{-2})$ 
\cite{Minwalla:1999px}. 
In this case when $\tilde{k}\ll \Lambda_{UV}^{-1}$
(i.e. when we are below the IR cut-off) we would have 
\begin{equation}
\Pi_{\mu\nu} \approx \Pi_1(k^2,\Lambda_{UV}^{-2}) \, \left( k^2 g_{\mu\nu} 
- k_\mu k_\nu \right)
+ \Pi_{2} (k^2, \Lambda_{UV}^{-2})\,\Lambda_{UV}^{4} \tilde{k}_{\mu}\tilde{k}_{\nu}
\,,
\end{equation}
and normal Wilsonian behaviour would be restored, with the couplings
matching those at the UV cut-off scale. Of course there is no reason to
suppose that such a cut-off in any way resembles what actually happens
in string theory, and to discuss the nature of the theory below
$\Lambda_{IR}$ requires full knowledge of the real UV completion. 
\if 
Indeed in
general there are infinitely many different UV complete theories
describing the same effective noncommutative physics in the intermediate
range, and correspondingly infinitely many theories below
$\Lambda_{IR}$. 
\fi
{\em What then are our general expectations for physics below
$\Lambda_{IR}$? Does it correspond to an effective field theory?
If so, what happens to the Lorentz violating divergences in the IR?}

These are the precise questions we would like to explore, using a 
framework in which 
the noncommutative gauge theory is realized
as a low-energy effective theory on D-branes 
\cite{Connes:1997cr,Douglas:1997fm,Chu:1998qz,Seiberg:1999vs}.
Our arguments are based on the two point function as calculated on D-branes
in the background of a non-zero $B$-field \cite{Bilal:2000bk,Chu:2000wp,Gomis:2000bn,Liu:2000ad}. 
In such a theory, taking the
zero slope limit in a particular way \cite{Seiberg:1999vs}
($\alpha'\rightarrow 0$ with $g_{\mu\nu}\sim \alpha'^2$) yields a
noncommutative field theory in which the role of the noncommutativity
parameter is played by the gauge invariant Born-Infeld field strength:
indeed in this limit the open string metric and the noncommutativity 
parameter are given by \cite{Seiberg:1999vs}
\begin{equation}
G^{\mu \nu} = \left(\frac{1}{g-F}g \frac{1}{g+F}\right)^{\mu\nu} \end{equation}
with $F_{\mu \nu}=2\pi \ap B_{\mu \nu}$, $B_{\mu\nu}$ being the (magnetic) field strength, and \begin{equation}
\theta^{\mu\nu} = -2\pi \ap \left( \frac{1}{g-F} F \frac{1}{g+F} \right)^{\mu\nu}
\end{equation}
respectively (we will henceforth restrict
ourselves to noncommutativity in the space directions which we will 
label $ij$). The theory at finite $\alpha'$ provides a convenient UV
completion of the noncommutative gauge theory. 
The UV "cut-off" acquires a physical meaning: it is
the scale above which the 
noncommutative field theory description is invalid and 
string modes become accessible,
and is of order 
\beq
\Lambda_{UV} =1/\sqrt{\alpha'}.
\eeq
The IR "cut-off" is accordingly given by 
\beq
\Lambda_{IR}= \sqrt{\alpha'} {M^2_{NC}},
\eeq
and, likewise, physics below this scale is best understood by performing 
a string calculation.  
We will rather loosely continue referring to the scale $\Lambda_{IR}$ as the 
IR cut-off although of course we are chiefly interested in exploring the effective theory below it. 

What we will show in this paper is that {\em the one-loop 
effective theory in the $k\rightarrow 0$ limit (including any 
threshold contributions) 
is the same as the commutative $\theta=0$ theory}, 
and in particular there are no IR divergences. 
Below $\Lambda_{IR}$, physics differs from the $\theta=0$ physics only by nonsingular 
residual  effects that are calculable in any specific model, and we will estimate their 
magnitude. 
In addition we point out that the two point function of the graviton 
also gets stringy contributions at one-loop which can modify gravity right down 
to $\Lambda_{IR}$: if 
for example  $M_{NC}\sim 1$TeV and $M_{s}\sim M_{Pl}$,
then gravity is modified at a $mm$ 
even when there are no large extra dimensions.
This is an effect equivalent to the one described for the gauge theory
however there is no simple effective field theory description and it is difficult 
to understand in terms of ``planar'' and ``nonplanar''.
 
The rest of the paper is organized as follows. In the next section, we
will discuss and determine the general form of UV/IR mixing in 
noncommutative field theory which is embedded in string theory. 
In section 3 and 4, the mentioned general characteristics of UV/IR
mixing will be justified with explicit amplitude calculations based on
bosonic and superstring models. In section 5, we will analyse how the graviton 
two point function is modified by noncommutativity.
In section 6,  we will discuss how
noncommutativity in string theory may lead to a modification in the IR
property of gravity. We will also discuss its phenomenological implications.

\section{General remarks on UV/IR mixing in string theory} 

Assuming that the string theory amplitudes are finite (as issue to which
we return in due course), it is natural that the IR singularities should be
cured in much the same way as UV singularities are, since they are
intimately related: they are essentially the same singularities. It is
also natural that string theory should cure discontinuities afflicting
the field theory; we certainly expect a string amplitude calculated at
non-zero ${F}$, which is after all a rather mild background, to tend
continuously to the one calculated at ${F} =0$. What is more striking is
that in a nonsupersymmetric theory the Lorentz violating $\Pi_2$ term
also tends to zero as $ \tilde{k}^2 /\alpha'$ below the IR cut-off,
reminiscent of the field theory behaviour with the naive Schwinger
cut-off. 

Consider nonplanar annulus amplitudes in bosonic string theory on a D$p$-brane. 
As we shall see, the general structure of a one loop diagram can be 
very heuristically 
written as 
\begin{equation}
A_{NP}\sim \int \frac{dt}{t} t^{-\frac{(p+1)}{2}} e^{ - \tilde{k}^2/4t} \, f(t).
\end{equation} 
The function $f(t)$ includes kinematic factors as well as sums over 
all the open string states in the loop. The integration parameter $t$ 
is the parameter describing the annulus. In the field theory limit
$\alpha'\rightarrow 0$  
we recover the expected nonplanar field theory contribution, with $t$ playing the 
role of a Schwinger parameter. In addition all but the massless open strings (and 
in this case the tachyon whose contribution we discard) do not 
contribute in this limit. In the present discussion we are of course 
not interested in taking the 
field theory limit but will instead 
keep  $\alpha'$ finite. The crucial feature of the
amplitudes governing the IR behaviour is that 
the nonplanar integrands always come with a factor $e^{ - \tilde{k}^2/4t}$
irrespective of whether we are above or below  $\Lambda_{IR}$. When 
$\tilde{k}^2 \gg \alpha' $ the integrand is killed everywhere in the stringy 
region $t < \alpha'$ 
and the amplitude is close to the field theoretical result. Indeed one 
may make a large $t$ expansion rendering the amplitude identical to the 
field theoretical one. On the 
other hand in the area of most interest below the IR cut-off we have 
$\tilde{k}^2 \ll \alpha' $ and hence stringy $t < \alpha'$ regions also
contribute to the integral. If the integrand is finite  
and free of singularities then in the limit as $k\rightarrow 0$ 
the amplitudes clearly tend continuously to their commutative equivalents.
Thus the finiteness of the string amplitudes immediately guarantees 
that the $k\rightarrow 0$ limits and the $\theta\rightarrow 0$ limits
give the same physics. 
Moreover in this limit we may expand the  $e^{ - \tilde{k}^2/4t}$
factor inside the integral. The nett result is that far below the IR cut-off 
one-loop amplitudes may be written as 
\begin{equation}
A(\theta,k)\sim A(0,k) (1 + \lambda \frac{\tilde{k}^2}{\alpha'} +\ldots) ,
\end{equation}
where $\lambda $ is a factor including loop suppression and gauge
couplings and the second piece is the leading term in the small
$\tilde{k}^2/\alpha'$ expansion of the exponential factor. Note that the
$A(0,k)$ prefactor includes the usual one-loop contributions of the
commutative theory and hence all stringy threshold corrections. Thus
although various compactification scenarios may result in vastly
different threshold corrections, the leading effect of non-zero $B$
field will always be of this form. (Extension to $N$-point amplitudes is
trivial.) 

Based on this generic expression for the amplitudes, phenomenology below
$\Lambda_{IR}$ takes on a characteristic form. First from the low energy
point of view the net effect of the non-zero $B$ field is simply to
take the non-planar contribution to thresholds of gauge couplings and
move them down to the IR cut-off, inserting between $\Lambda_{IR}$ and
$\Lambda_{UV}$ a region approximating conventional noncommutative field
theory. Below $\Lambda_{IR}$, the leading deviation from the commutative
theory (including all its stringy thresholds) has a factor
$\tilde{k}^2/\alpha' $, with the dimensionality being made up by powers
of $\alpha'$. 

Thus for example the $\Pi_2 $ term is of the form 
\beq \Pi_{2} (k^2, \tilde{k}^2) \sim 
 \lambda (\tilde{k}^2 \alpha')^2  \eeq
in a nonsupersymmetric theory and 
\beq \Pi_{2} (k^2, \tilde{k}^2) \sim \lambda (\tilde{k}^2 \alpha')^2
 \frac{M_{SUSY}^2}{\ap}    \eeq
in a theory with supersymmetry softly-broken at a scale $M_{SUSY}$. 
(Note that the factor of $ (\tilde{k}^2 \alpha')^2 $ is simply to undo 
the power of 
$ \tilde{k}^{-4}$ in the above definition of $\Pi_2$.) 
This introduces a birefringence into the trace-U(1) photon, a
polarization dependent velocity shift of order 
\beq
\Delta v \sim c \frac{\lambda M_{SUSY}^2 M_s^2}{M_{NC}^4}.
\eeq
This effect is much milder than the naive expectation and can be made
phenomenologically acceptable with a large $M_{NC}$ even if the 
physical photon is predominantly made of trace U(1) photon as described in 
ref.\cite{Abel:2006sp}. The model dependent issue here which we will 
expand upon in the 
following sections is the coefficient $\lambda $ 
which encapsulates the strength of the one-loop contributions 
(i.e. threshold corrections to couplings) relative to the tree level ones. 

If the physical photon is decoupled from the trace U(1) photon 
(see for example \cite{Chu:2001fe} where 
the trace U(1) photon becomes weakly coupled in the IR
and forms part of a hidden sector to break and mediate supersymmetry), 
then there can be interesting implications for gravity. 
Consider a theory where the physically
observed Planck scale  
receives significant one-loop threshold corrections from the open string sector. 
This contribution can be computed from
the two point function of the gravitons with the open string modes running in the 
loop. To gain some more intuition on what effects of
noncommutativity might be, we turn to an effective field theory
description. A reasonable (but, as it turns out, incorrect) 
guess for the effective field theory coupling the open string modes to the graviton
is a lagrangian of the form 
\begin{equation}
{\cal L} = \int d^4 x \sqrt{-g}g^{\mu\mu'} g^{\nu\nu'} 
\frac{F_{\mu\nu}* F_{\mu'\nu'}}{4g^2 } ,
\label{Naivegrav}\end{equation}
where there is, note, no star product between the "closed string 
metric" or in its determinant. The desired contribution can be computed from
the two point function of the gravitons with the gauge bosons running in the 
loop, and thus our effective field theory above 
would generate "planar" and "non-planar" 
diagrams exactly as in the pure gauge case, the crucial point being the 
presence of a Moyal phase coming from the vertices. 
Thus one might expect that in string theory, turning on a $B$-field
would separate planar 
and non-planar contributions to the graviton two point function, 
in much the same way as for the photon.  
Thanks to UV/IR mixing the nonplanar contributions would change all the way 
down to $\Lambda_{IR}$ below which they would asymptote to the values of the 
commutative theory.  There, the leading deviation in Planck's constant 
from that of the purely commutative theory  
should be precisely as described above for the gauge couplings. 
As we will see in the section 5 
the true picture is actually more subtle than this 
\footnote{In the field theory 
limit, an effective vertex involving a graviton and two photons exists (and 
indeed we compute it), 
but there is no such simple Lagrangian from which it could be derived.
}.  Nevertheless the effect 
we described persists; namely that subleading $\tilde{k}^2/\alpha'$ 
suppressed corrections in the two point function of the graviton 
lead to a modification of gravity at energy scales higher than $\Lambda_{IR}$.

\section{UV/IR mixing in the bosonic string}

We will first look at the 2-point function
on the annulus for pure QED, equivalent to the noncommutative Yang-Mills
action \beq
S=-\int\frac{1}{4}F_{\mu\nu}*F^{\mu\nu}\eeq
The contributions to the 2-point amplitudes on D$p$-branes in a noncompact
26-dimensional volume requires open string vertex operators
\begin{equation}
V = g_{D_p} \epsilon_{\mu} \partial X^{\mu} e^{ik\cdot x}.
\end{equation}
which have been appropriately normalised ($g_{D_p}^2 = (2\pi)^{p-2} g_c (\alpha^{\prime})^{\frac{p-3}{2}}$). This gives the amplitude
\begin{eqnarray}
A_{2}(k,-k) & = & -2\ap g_{D_p}^{2}V_{p}\int_{0}^{\infty}\d t\,(8\pi^{2}\alpha't)^{-\frac{(p+1)}{2}}\eta(it)^{-24}\times \nonumber \\
 &  & \,\,\int_{0}^{t}dx\left.e^{-2\alpha'k.G(x,x').k}
\left(\varepsilon_{1}.G_{xx'}.\varepsilon_{2}-
2\alpha'(\varepsilon_{1}.G_{x}.k)(\varepsilon_{2}.G_{x'}.k)\right)
\right|_{x'=0}.
\label{Bosmaster}\end{eqnarray}
Here $x, t$ play the role of \emph{dimensionless} Feynman and 
Schwinger parameters respectively. At this point, we should comment 
that throughout this paper we shall take the fundamental domain of the 
annulus to be $[0,1/2]\times[0,it]$.

Note that we write the measure with integration over all of the vertices, and then use the 
annulus' translation invariance to fix one vertex, including a volume factor of $t$. 
The one-loop Green's functions required
depend on whether the diagram is planar or nonplanar, and are given
by 
\cite{Bilal:2000bk,Gomis:2000bn,LiuM}
\begin{equation}
G^{\alpha \beta} (x, x^{\prime}) = I_0 \delta^{\alpha \beta} + 
J \frac{(\theta^2)^{\alpha \beta}}{\alpha^{\prime 2}} + 
K \frac{\theta^{\alpha \beta}}{\alpha^{\prime}},
\end{equation}
where, for the planar case,
\begin{equation}
I_0^{P} (x - x^{\prime}) = \log |t \frac{\theta_1 (\frac{x-x^{\prime}}{t},\frac{i}{t})}{\eta^3 (i/t)}| ,  \qquad 
J^P = 0 , \qquad
K^P (x - x^{\prime}) = -\frac{i}{4} \epsilon (x-x^{\prime}), 
\label{PropsPlanar}\end{equation}
and for the nonplanar case,
\begin{equation}
I_0^{NP} (x - x^{\prime}) = \log t \frac{\theta_4 (\frac{x-x^{\prime}}{t},\frac{i}{t})}{\eta^3 (i/t)} , \qquad
J^{NP} = \frac{-1}{8\pi t} , \qquad 
K^{NP} (x + x^{\prime})= \pm \frac{\pi}{t} (x+x^{\prime}), 
\label{PropsNonPlanar}\end{equation}
where the $+ (-)$ in $K^{NP}$ applies for the outer (inner) boundary. 
The feature of these expressions which ensures the regularization
of the nonplanar diagram is the contraction $k.G.k$ appearing in
the exponent of the integrand. We find \begin{eqnarray}
-2\alpha'k.G^{P}.k & = & -2\alpha'k^{2}I_{0}^{P}, \\
-2\alpha'k.G^{NP}.k & = & -2\alpha'k^{2}I_{0}^{NP}
-\frac{\tilde{k}^{2}}{8\pi\alpha't}.
\end{eqnarray}

Having established the Green's functions, we can perform the integration by parts in equation (\ref{Bosmaster}) to extract the kinematics. 
Defining (for ease of notation) $\hat{\Pi}_2=\Pi_2 \tilde{k}^{-4}$, we find 
\begin{eqnarray}
\label{amp}
A_2^P &=&  \Pi_1^P (k^2, 0) [ (\epsilon_1 \cdot \epsilon_2) k^2 - 
(\epsilon_1 \cdot k) (\epsilon_2 \cdot k)],  \\
A_2^{NP} &=& \Pi_1^{NP} (k^2, \tilde{k}^2) 
[ (\epsilon_1 \cdot \epsilon_2) k^2 - (\epsilon_1 \cdot k) 
(\epsilon_2 \cdot k)] + \hat{\Pi}_2 (k^2,\tilde{k}^2) 
[ (\epsilon_1 \cdot \tilde{k}) (\epsilon_2 \cdot \tilde{k})], \nonumber
\end{eqnarray}
where the standard gauge running is given by the formula
\begin{equation}
\Pi_1 = 4\alpha^{\prime 2} g_{D_p}^{2}\int_{0}^{\infty}\d t\,
Z(t)
e^{-\frac{\kt^2}{4\ap t} \times a} 
\int_0^t \d x \ e^{-2\ap k^2 I_0}(\dot{I}_0)^2 ,
\label{Pi1B}\end{equation}
while the Lorentz-violating piece is given by
\begin{equation}
\label{Pi2B}\hat{\Pi}_2 = 4\pi^2 g_{D_p}^{2}\int_{0}^{\infty}
\frac{dt}{t^2}
Z(t)
e^{-\frac{\kt^2}{4\ap t}}  \int_0^t \d x \ e^{-2\ap k^2 I_0^{NP}}.
\end{equation}
In eqn.(\ref{Pi1B}) $a=0$ or 1 for the planar or nonplanar case respectively. The term $Z(t)$ is the partition function for the model, which we shall take throughout to be
\begin{equation}
Z(t) = (8\pi^{2}\alpha' t)^{-\frac{(p+1)}{2}}\eta(it)^{d-24}.
\end{equation}
The parameter $d$ is inserted to remove adjoint scalars from the theory: there are $p-1$ physical polarisations of the photon, and the remaining $25-p$ modes are scalars, so we can interpret the parameter $d$ as removing $d$ of these modes. This is performed either by considering string theory in $26-d$ dimensions \cite{DiVecchia:1996uq,Bilal:2000bk}, or taking the spacetime to be, for example, $\mathbb{R}_{p+1} \times \mathbb{R}_{25-p-d} \times \mathbb{R}_d/Z_N$ with the $p$-brane at a singularity such that the orbifold projection removes the scalars \cite{Gomis:2000bn}. In this way, we can alternatively consider $d$ as modelling the effect of compactified dimensions, and hence we shall refer to $d$ ``compact dimensions'' throughout.

The forms (\ref{amp})-(\ref{Pi2B}) for the 1-loop amplitudes were already discussed in ref.\cite{Bilal:2000bk} in the field theory limit.
However it is important to note now that we have not taken any field theory 
limit and yet the $\tilde{k}$
dependence is already entirely contained within the $\tilde{k}^{2}$
term: the sole effect of noncommutativity is to truncate the Schwinger integration to $2\pi t>\frac{\tilde{k}^{2}}{\alpha'}$,
even in the full string expression. 

Thus there are two regimes that we will consider. The first is the
regime where $\tilde{k}^{2}/\alpha'\gg1$. In this case the Schwinger
integral is truncated to the region $2\pi t\gg1$ and the integral is
well approximated by the $t\rightarrow\infty$ limit. The second regime
is where $\tilde{k}^{2}/\alpha'\ll1$. In this case much of the Schwinger
integral is over the region where $2\pi t\ll1$ and one expects the
$t\rightarrow\infty$ limit to be a poor approximation. In this limit
a good approximation to the integral requires a modular transformation
of the $\vartheta$ and $\eta$ functions to the closed string channel.
It is natural to think of $\alpha$' as playing the role of the UV
cut-off to the field theory, $\alpha'\equiv\Lambda_{UV}^{-2}$, 
and then this regime corresponds
precisely to \beq
k\ll\frac{M_{NC}^{2}}{\Lambda_{UV}}\equiv\Lambda_{IR},\eeq
i.e. the region in the deep IR where the field theory computation
breaks down. 

Indeed the integral will become sensitive to the global structure
of the compactified dimensions since the $t\rightarrow0$ UV end of
it corresponds to closed string modes in the deep IR. 
Note that \cite{Armoni:2001uw,Armoni:2003va} studied the connection between IR poles and closed
string tachyons; we shall neglect these as we are interested in extracting
results relevant to consistent theories. 
There may be other thresholds as well as the string scale where for example winding
modes of the compact dimensions start to contribute in the integral.
In order for there to be an effective field theory description below
$\Lambda_{IR}$ these effects should add contributions independent
of $p$. In order to incorporate these effects one can divide the Schwinger 
integral into regions $t\in[0,1]$ and $t\in[1,\infty]$
where the two approximations are valid. 

\subsection{Brief review of planar diagrams }
The methods for obtaining the low energy behaviour of string
diagrams in order to derive the effective four-dimensional field
theories have been well covered elsewhere
\cite{DiVecchia:1996uq,DiVecchia:1996kf}. 
Since the integrals do not
contain any evidence of the non-commutativity, as can be seen from the
Green's functions, the only difference from the $B=0$ calculation
in this case is a phase dependence on the ordering of the vertices.
The reader is referred to
the Appendix for details of the $t\rightarrow\infty$ limit for planar
diagrams, which
reviews some of the basic techniques that we will be using. 
We shall consider $d$ dimensions transverse to the brane to be compactified with a radius close to or at the string scale, although we shall pay special attention to the case $d=0$. The contributions to $\Pi_1$ and $\hat{\Pi}_2$ in this limit are given in equations (\ref{Pi1PD}) and (\ref{Pi1PD0}).

The $t\rightarrow0$ limit of the planar diagrams is
the UV contribution, that is $t\in[0,1]$ indicating
momenta much higher than the string scale. It is well known  
as the planar contribution to the string threshold 
correction, however we present it here in order to emphasise
the way that
string theory is thought to render such contributions finite. 
Let us compute the $t\rightarrow 0$ contribution to 
the two point functions for a D$p$-brane in 
26 noncompact dimensions. 
We modular transform the expressions to give for the partition function
\begin{equation}
(8\pi^{2}\alpha't)^{-\frac{(p+1)}{2}}t^{12}e^{\frac{2\pi}{t}}(1+24\, 
e^{-\frac{2\pi}{t}}+\ldots),
\end{equation}
where, since we are assuming no compact dimensions, 
there are no winding modes, and thus
\begin{multline}
\Pi_{1\  UV}^P \rightarrow \frac{g_{D_p}^{2}}{16\pi^2 (8\pi^{2}\alpha')^{\frac{(p-3)}{2}}} \int_{0}^{1}\d t\, t^{\frac{(21-p)}{2}} (2t)^{-2\alpha'k^{2}} \int_{0}^{1}dy\, \\
\bigg[ 24 |\cos \pi y|^2 |\sin \pi y|^{-2 -2\ap k^2} + 8 |\cos \pi y|^2 
|\sin \pi y|^{-2\ap k^2} \bigg].
\end{multline}
Here we cannot neglect the ``pole'' pieces, but perform the integral in terms of beta functions and analytically continue in the momentum, using
\begin{equation}
\int_0^1 \d y |\cos \pi y|^a |\sin \pi y|^b = \frac{2}{\pi} B(\frac{a+1}{2},\frac{b+1}{2})
\end{equation}
 to give the zero-momentum limit
\begin{equation}
\Pi_{1_{UV}}^{P}= \frac{5 g_{D_p}^{2}}{2\pi^2 (23-p) (8\pi^{2}\alpha')^{\frac{(p-3)}{2}}},\end{equation}
a threshold contribution to the gauge couplings which is finite when
$p<23$. 

Now, of course this computation is a cheat
because it assumes that
transverse space was noncompact. In a compact space, sooner or later
in the $t\rightarrow0$ limit we need to sum over winding sectors
in the measure of the integral. Once the winding sectors are included
the effective $p$ is $p\equiv25$ and the integral diverges. However
this divergence is resolved in a way that is at least qualitatively
well understood: the natural way to write $A_{2_{UV}}$ is with the
parameter $S=\frac{\alpha'^{2}}{T}$ which reveals the expression
to be in the $\alpha'k^{2}\ll$1 limit simply an IR pole due to a
massless closed string tadpole. Indeed in this limit and when $p=25$
the contribution from level $n$ is proportional to \beq
\int_{0}^{\alpha'}\frac{dT}{T^{2}}e^{-\frac{4\pi^{2}\alpha'}{T}(n-1)}=\int_{\alpha'}^{\infty}dS\, e^{-\frac{4\pi^{2}}{\alpha'}(n-1)\, S},\eeq
as appropriate for closed string 
states with $m^{2}=\frac{4\pi^{2}}{\alpha'}(n-1)$.
Such tadpoles are a signal that we are expanding around the wrong vacuum,
and the solution is to give a VEV to the relevant fields in order to 
remove them. In this way the background is modified by the presence of the 
tadpoles and the
nett effect is that systems with $>2$ codimensions (i.e. $p<23$)
are insensitive to the moduli of the transverse dimensions, whereas
those with 1 or 2 codimensions get threshold corrections that are
respectively linearly or logarithmically dependent on the size of
the transverse dimensions, but are still believed to be finite even
when supersymmetry is broken by the construction. In principle in
certain tachyon-free nonsupersymmetric cases one can resum
the tadpole contributions to the tree-level perturbation series to
achieve a finite result. The precise details are rather subtle and
beyond the scope of this paper, and the reader is referred to 
refs.\cite{Fischler:1986tb,Antoniadis:1998ax,Dudas:2004nd} for more details.

\subsection{Non-planar diagrams in the $\tilde{k}^{2}/\alpha'\gg1$ limit}

We now turn to the non-planar diagrams.
Once we turn on the $B$-field, 
the presence of the $e^{-\frac{\tilde{k}^{2}}{8\pi\alpha't}}$ regulating
factor will cause the celebrated 
UV/IR mixing. We may treat the UV and IR contribution at the same
time in the two limits $\tilde{k}^{2}/\alpha'\ll1$ 
and $\tilde{k}^{2}/\alpha'\gg1$. 
Consider the second of these limits. 
The integrand is killed in the region where $t\ll\tilde{k}^{2}/\alpha'$
and hence we may always use the large $t$ limit of the integrand. 
We obtain
\begin{eqnarray}
\Pi_{1} & = & \frac{ g_{D_p}^{2}}{(4\pi)^{\frac{p+1}{2}}} \int_{2\pi \ap}^{\infty}dT\, T^{-\frac{(p-1)}{2}}\int_{0}^{1}dy\,\left[(24-d)(1-2y)^{2}-8\right]e^{-T k^{2}(y-y^{2})-\frac{\tilde{k}^{2}}{4T}}\nonumber \\
 & \rightarrow & \frac{ g_{D_p}^{2}}{(4\pi)^{\frac{p+1}{2}}}  \int_0^1 \d y \ 
\left[ (24-d)(1-2y)^{2}-8 \right] \bigg[ \frac{4k^2 y(1-y)}
{\kt^2}\bigg]^{(\frac{p-3}{4})} K_{\frac{p-3}{2}}(\sqrt{y(1-y)k^2
\kt^2})
\nonumber \\
 & \approx & \left\{ \begin{array}{rc} \frac{d}{3} g_{D_p}^{2} 
(4\pi)^{-\frac{p+1}{2}} \ln k^2 \kt^2,  & p=3 ,
\\ \frac{d}{3} g_{D_p}^{2} 
(4\pi)^{-\frac{p+1}{2}} 2^{p-5}\Gamma(\frac{p-3}{2})|\tilde{k}|^{3-p}, & p > 3, 
\end{array} \right.
\end{eqnarray}
where in the last step we assumed $|k||\tilde{k}|\ll1$ or in other
words momenta $|k|\ll M_{NC}$. In the case $p=3$ this gives the
same logarithmic running to a free field theory in the IR observed
in the field theory. When $p>3$ we find power law running in the
IR as described in \cite{Abel:2005rh}. The Lorentz-violating
term $\hat{\Pi}_{2}$ is given by \begin{eqnarray}
\hat{\Pi}_{2} & = & \frac{ g_{D_p}^{2}}{(4\pi)^{\frac{p+1}{2}}} \int_{2\pi\ap}^{\infty}dT\, T^{-\frac{(p+3)}{2}}\int_{0}^{1}dy\, (24-d) e^{-T k^{2}(y-y^{2})-\frac{\tilde{k}^{2}}{4T}} \nn \\
 & \approx & (24-d) \frac{ g_{D_p}^{2}}{(4\pi)^{\frac{p+1}{2}}} 2^{p-1}\Gamma(\frac{p+1}{2})|\tilde{k}|^{-(p+1)} \end{eqnarray}
and shows a similar power law behaviour in the IR. For $p=3$ and 
$d=22$ we reproduce the result of \cite{Bilal:2000bk}. This behaviour
is entirely in line with what one would expect from the field theory.

\subsection{Non-planar diagrams in the $\tilde{k}^{2}/\alpha'\ll1$ limit}

In this limit one expects to find behaviour differing from noncommutative
field theory. We now have to split the integral into two halves, $t>1$
and $t<1$. The first IR part is treated similarly to
the previous section, except in this case we simply set $\kt = 0$ in the integrand when 
we consider $\ap \rightarrow 0$, and should thus obtain the same results as
in the planar case; it is straightforward to show that for $p>3$ \begin{eqnarray}
\Pi_{1_{IR}}^{NP} & \approx &  \frac{d}{3}
\frac{g_{D_p}^{2}}{(4\pi)^{\frac{p+1}{2}}}  
\frac{2}{(p-3)} (2\pi\alpha')^{\frac{3-p}{2}}, \nn \\
\hat{\Pi}_{2_{IR}}^{NP} & \approx & (24-d)
\frac{g_{D_p}^{2}}{(4\pi)^{\frac{p+1}{2}}}  
\frac{2}{(p+1)} (2\pi\alpha')^{\frac{-(p+1)}{2}}\, .\end{eqnarray}
The contributions are roughly constant, and
equal to those of the $\tilde{k}^{2}/\alpha'\gg1$
limit when $\tilde{k}^{2}=4\alpha'$.

The second, UV, contribution for $t<1$ is the most interesting,
as it is this contribution which in field theory gives IR poles. We now 
modular-transform the expressions, and expand in powers of $e^{-\frac{2\pi}{t}}$. For no
compact dimensions, we have
\begin{equation}
\Pi_{1\ UV}^{NP} \rightarrow \frac{g_{D_p}^{2}}
{16\pi^2 (8\pi^{2}\alpha')^{\frac{(p-3)}{2}}} \int_{0}^{1}\d t\, 
t^{\frac{(21-p)}{2}} (2t)^{-2\alpha'k^{2}} e^{-\frac{\kt^2}{8\pi\ap t}} 
e^{-\frac{\pi\ap k^2}{2t}}\int_{0}^{1}dy\, \sin^2 2 \pi y.
\end{equation}
Note that for $\frac{\tilde{k}^{2}}{\alpha'}\rightarrow0$ the integration
is finite and the integral goes continuously to that of the 
commutative 
contribution, i.e. we have \beq
\Pi_{1}^{NP}(\theta)=\Pi_{1}^{NP}(\theta=0)\bigg(1+{\cal O}( \frac{\tilde{k}^{2}}{\alpha'})\bigg),\eeq
as promised in the Introduction; in other words, at momenta \emph{$k\ll\Lambda_{IR}$ the Wilsonian
gauge couplings return to the values they would have had for a completely
commutative theory with the same gauge group}. Note that this statement
is expected to be true even when $p\geq 23$ and in compact spaces for the following reason. In the finite examples we have seen, the effect of string theory is clearly 
to allow the limit $\tilde{k}\rightarrow 0$ to be taken continuously,
and to give the same physics as $\theta=0$. If this 
is true of any consistent UV completion, then it seems reasonable
to assume that what's good for the planar diagrams is good for the 
nonplanar ones. In other words, if the diagrams are formally divergent, 
continuity demands that the vacuum shifts which remove the 
UV divergences (i.e. closed string tadpoles) in the $B=0$ theory 
should do so in the $B\neq 0$ theory as well,
upto ${\cal{O}}(\tilde{k}^2/\alpha')$ corrections. 
Note that this is true even though the non-planar diagrams do not 
factorise onto disks in the closed string channel; IR singularities
arise from divergences in the partition function which are regulated by the 
$e^{-\frac{\kt^2}{8\pi\ap t}} e^{-\frac{\pi\ap k^2}{2t}}$ term, and so when these divergences are
cancelled, so are the IR poles. 

This reasoning leads one to expect that the $\hat{\Pi}_{2}$ term is regulated, since it should tend to zero as $\theta\rightarrow 0$.
Let us check this by computing the final contribution which is \begin{eqnarray}
\hat{\Pi}_{2_{UV}}^{NP} & \rightarrow & \frac{24 g_{D_p}^{2}}{(8\pi^{2}\alpha')^{\frac{(p+1)}{2}}}\int_{0}^{1}\frac{\d t}{t^{4}}\, t^{\frac{(25-p)}{2}} t^{-2\alpha'k^{2}}e^{-\frac{\kt^2}{8\pi\ap t}} e^{-\frac{\alpha'k^{2}\pi}{2t}} \int_{0}^{1}dy\,\nn \\
 & \approx & \frac{24}{19-p}\frac{g_{D_p}^{2}}
{(8\pi^{2}\alpha')^{\frac{(p+1)}{2}}}. 
\end{eqnarray}

\section{Supersymmetric models}

To include the effects of worldsheet fermions we require the fermionic propagators 
\cite{Liu:2000ad}:
\begin{equation}
\langle \psi^{\alpha} (z_1) \psi^{\beta} (z_2) \rangle_{\nu} = 
G^{\alpha \beta} \frac{\theta_{\nu} (z_1 - z_2) \theta_1^{\prime} (0)}
{\theta_{\nu} (0) \theta_1 (z_1 - z_2)} ,
\end{equation}
where the index $\nu$ specifies the spin structure, which must be 
summed over in the full amplitude. The above differs from the usual 
boundary fermion propagators purely by the replacement of the metric 
by the open string metric, but when we perform the rescaling of the 
external momenta and polarizations \cite{Bilal:2000bk} it is transformed 
back to the standard propagators:
\begin{eqnarray}
\langle \psi^{\alpha} (z_1) \psi^{\beta} (z_2) \rangle_{\nu} &\rightarrow& 
\delta^{\alpha \beta} \frac{\theta_{\nu} (z_1 - z_2) \theta_1^{\prime} (0)}
{\theta_{\nu} (0) \theta_1 (z_1 - z_2)} \nonumber \\
&\equiv& \delta^{\alpha \beta} G^{\psi}_{\nu} (z_1 - z_2),
\end{eqnarray}
which we shall use from now on. 

We wish to calculate the one-loop amplitude for two spacetime bosons with an 
arbitrary amount of supersymmetry in the loop, which is defined by the compact 
dimensions - and thus only affects the amplitude via the partition function. 
The vertex operators are
\begin{equation}
V^0 = g_{Dp} \epsilon_{\mu} (i\dot{X}^{\mu} + 2 \alpha^{\prime} k \cdot \psi 
\psi^{\mu} ) e^{i k \cdot X}\, ,
\end{equation}
and the resulting amplitude gives 
\begin{equation}
\Pi_1 = 4 g_{Dp}^2 (\ap)^2 \int_0^{\infty} \mathrm{d} t \sum_{\nu} \ e^{2\ap \kt^2 J} Z_\nu (t) \int_{0}^{t} \mathrm{d} x e^{-2\alpha^{\prime} k^2 I_0}  \bigg[(\dot{I}_0)^2 - (G^{\psi}_{\nu} (z(x))^2 \bigg]
\end{equation}
and
\begin{equation}
\hat{\Pi}_2 = 4\pi^2 g_{Dp}^2  \int_0^{\infty} \mathrm{d} t \sum_{\nu} 
\ Z_\nu (t) e^{ \frac{\kt^2}{4\ap t}} \int_{0}^{t} \mathrm{d} x 
e^{-2\alpha^{\prime} k^2 I_0} ,
\end{equation}
where $Z_{\nu}(t)$ is the partition function for the theory, and
\begin{eqnarray}
z^{P} &=& ix, \nonumber \\
z^{NP} &=& ix - 1/2.
\end{eqnarray}
Thus the spacetime fermionic component does not contribute to the
Lorentz-violating term, since the kinematics for it are just the
standard commutative gauge pieces. The Lorentz-violating term thus
derives from bosonic correlator exactly as in the bosonic string, the
only difference being the partition function. Of course, if there is
any supersymmetry then this term will vanish, as we expect, and the
remaining Lorentz-preserving term can be calculated from the off-shell
continuation of the fermionic piece. For $N \ge 1$ SUSY, $\Pi_1$ can
be simplified using the identity
\begin{equation}
(G^{\psi}_{\nu}(z))^2 = \frac{ \theta_{\nu}^{\prime \prime} (0)}{\theta_{\nu} (0)} 
- \partial^2 \log \theta_1 (z)
\label{fermpropsq}\end{equation}
to give
\begin{equation}
\Pi_1 = 4 g_{Dp}^2 (\ap)^2 \int_0^{\infty} \mathrm{d} t 
\sum_{\nu} \ e^{2\ap \kt^2 J} Z_\nu (t) 
\frac{ \theta_{\nu}^{\prime \prime} (0)}{\theta_{\nu} (0)} 
\int_{0}^{t} \mathrm{d} x e^{-2\alpha^{\prime} k^2 I_0}  .
\end{equation}
Again this is essentially the usual expression for computing threshold corrections,
but with an exponential factor inserted for non-planar diagrams. 

To summarize the results of this and the previous two sections,
as advertised in the Introduction, both bosonic and supersymmetric theories 
are found to tend continuously to the 
$B=0$ theory as $k\rightarrow 0$. In particular the couplings freeze out 
below $\Lambda_{IR}$ and the entire region above $\Lambda_{IR}$ can 
now be consistently integrated out in the usual Wilsonian manner.  
The phenomenological footprint of the non-zero $B$ field is then 
in the dispersion relation of massless particles, and in particular 
a birefringence of the trace-U(1) photon, which gets
a polarization dependent velocity shift of order 
\beq
\Delta v \sim c \frac{M_{SUSY}^2 M_s^2}{M_{NC}^4}.
\eeq
Whether the EM photon feels this effect is a model 
dependent question.

\section{The two point function of the graviton}

We now turn to the effect of the non-zero $B$ field 
on gravity by focussing on the graviton two-point function.
In particular, consider the corrections to the Newtonian 
force law of gravity due to the coupling of the graviton to gauge fields at one-loop.
The momentum dependence of these corrections determines the running of Planck's constant,
and our experience with gauge couplings suggests that this also may 
be subject to UV/IR mixing.

In the naive extension of noncommutative field theory of eq. (\ref{Naivegrav}), 
the one loop contributions divide into planar and non-planar 
exactly as they do for the trace U(1) photon. However in string theory the 
relevant diagram is an annulus with two  
graviton (closed string) vertices on the interior of the 
world sheet, and so the only way that planar could be distinguished from 
nonplanar would be either for there to be some kind of radial ordering 
effect in the vertices, or for there to be a limit in which the 
major contribution to the diagrams came from when the vertices 
were on the edges of the annulus. Neither of these possibilities 
is true and so, even before making any computation,
it seems unlikely that there will be a simple field theory approximation
involving Moyal products. 
The field theory limit has been the subject of a recent
study in ref.\cite{Alvarez-Gaume:2006bn} where it was indeed
found to be a rather complicated issue. However for the present study we do not
need to derive the effective action (and indeed we don't): 
we will instead examine the modification of the Newtonian 
force law between matter (open string) fields on the brane, by looking at 
the two point function determined at the string theory level. 

By restricting our attention to the force law between matter fields, 
we are evading a significant technical difficulty, namely
that in a sense we have two metrics, one for open strings and one for closed. 
We wish to examine the momentum dependence of the gravitational 
force between {\em open strings} confined to magnetised $D$-branes.
In principle we ought to be doing this by factorizing a four point open string amplitude 
on the graviton two point function.
The relations for $G,g$ and $\theta$ imply
\begin{equation}
g^{\mu\nu}= G^{\mu\nu} - \frac{(\theta G \theta)^{\mu\nu}}{4\pi^2 (\ap)^2}.
\label{openclosed}\end{equation}
determining the coupling of the matter on the brane to
gravity. 
We must choose a coordinate system where the components of the metric
$g_{\mu\nu}$ are made small ($~\ap F_{\mu\nu}$) for the dimensions 
in which magnetic field is turned on, 
so that the noncommutativity tensor $\theta^{\mu\nu}$ can 
be tuned to the desired values. 
Then the relevant momentum scale for the amplitude is given by the 
Mandlestam variables running through the loop, determined from the 
external momenta {\em as 
contracted with the open string metric}. Importantly the closed
string metric is vastly different in the regimes of interest: 
for the exchange of a 
graviton with four-momentum $q_\mu$ between open strings, the Mandelstam 
variables correspond to scales of order
\begin{equation}q^2 \equiv q_{\mu} q_{\nu} G^{\mu \nu}; 
\end{equation}
but if we were interested in 
graviton exchange between external graviton states, it would be more appropriate to use  
\begin{equation}
q_{\mu} q_{\nu} g^{\mu \nu} = q^2 - \frac{\tilde{q}^2}{4\pi^2 (\ap)^2},
\end{equation}
which, by definition, for $q \sim \Lambda_{IR}$, would be of order $\sim M_{s}^2$. 
In the former case, as we are only dealing with graviton propagators the difference is immaterial
since we can always rescale the graviton states to absorb the difference, but the 
correct procedure for (or indeed physical meaning of) the latter is less clear. 
(For example we would probably want more information about the other contributions
in such a process coming from $B$ fields, and also more information about what the 
asymptotic states are -- i.e. the effective field theory.) Thus in calculating the graviton correlator, 
we shall decompose the square of the momentum in terms of the open string quantities, 
and consider $\kt^2/\ap \gg \ap k^2$, while $k^2 \kt^2 \ll 1$.

We shall restrict the discussion to exchanges between $D3$-brane states, for 
which (since there can be no orientifold planes) we need only consider the annulus for the 
induced gravity on the brane at one loop. A typical model for this scenario would be $D3/D7$-branes at a $C^6/\mathbb{Z}_N$ orbifold singularity \cite{Aldazabal:2000sa}, with a magnetic flux on the $1$ and $2$ directions. 
However, we shall keep the discussion as general as possible.
We proceed initially as in ref.\cite{Antoniadis:1996vw,Epple:2004ra} to extract the correction to $\frac{M_{Pl}^2}{16\pi}$, 
denoted $\delta$, by  
considering the following kinematic portion of the graviton two-point function:
\begin{equation}
\langle V_G (h^1,k) V_G (h^2, -k) \rangle \supset 
-\frac{\delta}{4} \, h_{\mu \nu}^1 h^2_{\lambda \rho} g^{\mu \lambda} k^{\nu} k^{\rho} \equiv A_{\delta}
\end{equation}
where the vertex operators are given by
\begin{equation}
V_G (h,k) = g_s^2 N_G \frac{2}{\ap} h_{\mu \nu} 
\big(\partial X^{\mu} (z) -  \frac{i\ap}{2}: k \cdot \psi \psi^{\mu} (z): 
\big)
\big(\bar{\partial} X^{\nu} (\bar{z}) - \frac{i\ap}{2} : 
k \cdot \tilde{\psi} \tilde{\psi}^{\nu} (\bar{z}):
\big) e^{i k \cdot X(z, \bar{z})}.
\end{equation}
In the open string channel the Green's function is given by 
modular tansforming the result of 
ref.\cite{Liu:2000ad}\footnote{Note that this choice of propagator differs slightly from those given elsewhere \cite{Gomis:2000bn,Bilal:2000bk}, but it was asserted in \cite{Liu:2000ad} that the additional terms are necessary to ensure periodicity and obediance to the equations of motion. They cause a discrepancy when the fields are taken to the boundary; (\ref{PropsPlanar}, \ref{PropsNonPlanar}) are not obtained from (\ref{PropsClosed}) . However, the closed string propagators only differ by linear terms in $J^C$ and $K^C$, plus the function $f$ which plays no essential role in amplitudes (merely ensuring that the derivatives of the logs in the antisymmetric portion contain no discontinuities). The reader can check that (\ref{TachCorr}) is unchanged by these, and since $I^C$ is identical for both versions, so are all the other results in this section.}:
\begin{equation}
\mathcal{G}^{\mu \nu}(w_1,w_2) = -\frac{\ap}{2} I^C G^{\mu \nu} + J^C 
\frac{(\theta^{\mu \alpha}G_{\alpha \beta} \theta^{\beta \nu})}{8\pi^2 \ap} 
- K^C \frac{\theta^{\mu \nu}}{2\pi},
\end{equation}
where \begin{eqnarray}
I^C &=&  \ln \left| \frac{\theta_1 (w_1 - w_2,it) 
\theta_1 (w_1 + \bar{w}_2,it)}{4\pi^2 \eta^6(it)} \right|^2 
- \frac{4\pi}{t} | \Im(w_1 - w_2) |^2 , \nonumber \\
J^C &=&  \ln \left| \frac{\theta_1 (w_1 - w_2,it)}{ \theta_1 (w_1 + \bar{w}_2,it)} 
\right|^2 - \frac{4\pi}{t} \bigg(| \Re(w_1 + \bar{w}_2) |^2 - \Re(w_1) 
- \Re(w_2) \bigg), \label{PropsClosed}\\
K^C &=& \ln \theta_1 (w_1 + \bar{w}_2,it) - \ln \theta_1 (\bar{w}_1 + w_2,it)- 
\frac{2\pi i}{t} \Im \left( (w_1 + \bar{w}_2 + 1/2)^2 \right) 
-2\pi i f(\Im(w_1 - w_2))\,  \nonumber
\end{eqnarray}
and where $f(x) \equiv - [x/t]$, $[y]$ denotes the closest integer to $y$. Thus the self-contraction terms, 
with normal-ordering and the $w_1 \rightarrow w_2$ limit performed, are
\begin{eqnarray}
C^{\mu \nu} (w, \bar{w}) = && -\bigg(\frac{\ap}{2} G^{\mu \nu} + 
\frac{(\theta^{\mu \alpha}G_{\alpha \beta} \theta^{\beta \nu})}{8\pi^2 \ap}\bigg) \ln \left| 
\frac{\theta_1 (w + \bar{w},it)}{2\pi \eta^3(it)} \right|^2 \nonumber \\
&& -\frac{(\theta^{\mu \alpha}G_{\alpha \beta} \theta^{\beta \nu})}{8\pi^2 \ap} \frac{8\pi}{t} (2\Re^2(w) - \Re(w)) \, .
\end{eqnarray}

The fermionic Green's functions are obtained from the torus functions using the doubling trick:
\begin{equation}
\psi^{\mu} (w) = \left\{ \begin{array}{lc} \psi^{\mu} (w) , & \Re(w) > 0 , \\ 
i (\frac{g+F}{g-F})^{\mu}_{\nu} \tilde{\psi}^{\nu} (-\bar{w}),  & \Re(w) < 0. 
\end{array} \right.
\end{equation}
We obtain
\begin{eqnarray}
\langle \psi^{\alpha} (z) \psi^{\beta} (w) \rangle_{\nu} &=& g^{\alpha \beta}
 G_{\nu}^{\psi} (z-w), \nonumber \\
\langle \tilde{\psi}^{\alpha} (\bar{z}) \tilde{\psi}^{\beta} (\bar{w}) 
\rangle_{\nu} &=& g^{\alpha \beta} G_{\nu}^{\psi} (\bar{z}-\bar{w}), \nonumber \\
\langle \psi^{\alpha} (z) \tilde{\psi}^{\beta} (\bar{w}) \rangle_{\nu} 
&=& -i \left(g^{\alpha \beta} + 2\frac{(\theta G \theta)^{\alpha \beta}}
{4 \pi^2 (\ap)^2} -2 \frac{\theta^{\alpha \beta}}{2\pi \ap} \right) 
G^{\psi}_{\nu} (z + \bar{w}).
\end{eqnarray}
As for the gauge bosons, the physical behaviour naturally splits into long distance 
 $\tilde{k}^2/\ap \ll 1$ and short distance  $\tilde{k}^2/\ap \gg 1$ regimes.
In the former, gravity will be
dominated by the low energy modes, for which the usual corrections to
Planck's constant apply. We can expand the amplitude as a power series in $k^2$ and $\kt^2$, and neglect the terms $O(k^2)$ relative to $O(\kt^2)$. In the short distance regime however such an expansion is no longer appropriate, but the amplitude still has terms with a prefactor of $\kt^2$ which we should consider dominating over those prefixed by $k^2$. In this way, we may consider the same correlators as being typical dominating terms in the amplitude for the non-zero $B$-field corrections to  both limits; one such term is
\begin{equation}
A \supset \int_0^{\infty} \d t \int \d^2 z \int \d^2 w \, - g_s^2 \frac{\ap}{2} N_G^2 \langle \partial X^{\mu} (z) \bar{\partial} X^{\lambda} (\bar{w}) e^{i k \cdot X(z,\bar{z})} e^{-i k \cdot X (w,\bar{w})} \rangle \langle k \cdot \tilde{\psi} \tilde{\psi}^{\nu} (\bar{z}) k \cdot \psi \psi^{\rho} (w) \rangle
\end{equation}
which has a leading contribution of the form 
\begin{equation}
\int_0^{\infty} \d t \int \d^2 z \int \d^2 w \,  g_s^2 N_G^2 
\frac{\tilde{k}^2}{4\pi^2 \ap} Z(t) (G^{\psi}_{\nu} )^2 k_a k_b 
\partial_z \mathcal{G}^{a \mu} \bar{\partial}_w \mathcal{G}^{b \lambda} 
\langle e^{i k \cdot X(z,\bar{z})} e^{-i k \cdot X (w,\bar{w})} 
\rangle \equiv L_{\delta} + ...\, .
\end{equation}
Here $L_{\delta}$ is the component of this term which contributes to
$A_{\delta}$, and we have included in the partition function the
Chan-Paton summation, which corresponds to summing over the Casimirs
of the representations of the gauge group. We shall leave a complete
analysis to future work, and consider the contribution from the corner
of the moduli space where $t > 1$. Here we can take the derivatives of
the Green's functions to be given by the leading order terms as $t
\rightarrow \infty$ - as for the gauge theory case, this is equivalent
to a field theory calculation, but it is more expedient to perform the
calculation from string theory. We find that the behaviour is
dominated by the correlator of the exponentials: this is given by
\begin{multline}
\langle e^{ik \cdot X (z_1)} e^{-ik \cdot X (z_2)}\rangle =  \left| \frac{\theta_1 (z - w,it)}{2\pi \eta^3(it)} \right|^{-2\ap k^2 - \frac{\kt^2}{4\pi^2 \ap}}\left| \frac{\theta_1 (z + \bar{w},it)}{2\pi \eta^3(it)} \right|^{-2\ap k^2 +  \frac{\kt^2}{4\pi^2 \ap}} \\
\left| \frac{\theta_1 (z + \bar{z},it)}{2\pi \eta^3(it)} \right|^{\ap k^2 + \frac{\kt^2}{8\pi^2 \ap}}\left| \frac{\theta_1 (w + \bar{w},it)}{2\pi \eta^3(it)} \right|^{\ap k^2 + \frac{\kt^2}{8\pi^2 \ap}} \\
\exp\left[\frac{4\pi\ap k^2}{t} |\Im(z - w)|^2\right]
\exp\left[\frac{-\kt^2}{\pi \ap t}|\Re(z - w)|^2\right]. 
\label{TachCorr}\end{multline}
Note that this correctly factorises onto the corresponding boundary amplitude. To take the field theory limit now, we write $T = \pi \ap t$, $y = \Im(z)/t$, use the translation invariance of the annulus to fix $\Im(w) = 0$, and write $\Re(z) = x, \Re(w) = x^{\prime}$, and insert the partition function and kinematic factors, with a sum over spin structures. Making use of the identity (\ref{fermpropsq}) and assuming $N\ge1$ supersymmetry, so that after multiplying by the partition function 
all spin-structure-independent terms vanish, we obtain the prefactor
\begin{equation}
F(T) \equiv (8\pi T)^{-2} \sum_{\nu} Z_{\nu} (\frac{T}{\pi \ap}) \frac{ \theta_{\nu}^{\prime \prime} (0)}{\theta_{\nu} (0)}  \, .
\end{equation}
We shall assume $F(T)$ has the behaviour
\begin{equation}
\lim_{T\rightarrow \infty} T^2 F(T) = \beta,
\end{equation}
where $\beta$ is a constant. 
If we now insert the factors from our ``typical'' contribution, we obtain 
\begin{multline}
L_{\delta}^{FT} =  g_s^2 N_G^2 \frac{\tilde{k}^2}{4\pi^2 \ap} \int_{\pi \ap}^{\infty} \d T \ T F(T) \int_0^1 \d y (1-2y)^2 e^{-4k^2 T y(1-y)} \\
\int_0^{1/2} \d x \int_0^{1/2} \d x^{\prime} e^{\frac{-\kt^2}{T}(x -
x^{\prime})^2} \left|\sin 2\pi x \sin 2\pi x^{\prime}\right|^{\ap k^2 + \frac{\kt^2}{8\pi^2 \ap}}. 
\end{multline}
As discussed, in contrast to a noncommutative field theory, 
there is no separation into planar and non-planar diagrams.

Since we are considering the regime $k^2 \tilde{k}^2 \ll 1$, we reorder the integration and use the leading behaviour of the Bessel function $K_0$ as in ref.\cite{Bilal:2000bk} to give
\begin{eqnarray}
L_{\delta}^{FT}&\approx &  g_s^2 N_G^2 \frac{\tilde{k}^2}{4\pi^2 \ap}
\frac{\beta}{6\pi} \log |k^2\kt^2| B^2\left(\frac{1}{2},
\frac{1}{2}+\frac{\ap k^2}{2} + \frac{\kt^2}{16\pi^2 \ap}\right) \nonumber \\
& \approx &  g_s^2 N_G^2  \frac{\tilde{k}^2}{4\pi^2 \ap} \frac{\beta}{6\pi} 
\frac{16 \pi^3 \ap}{\kt^2} \log |k^2\kt^2| 
\end{eqnarray}
and thus this contribution to the graviton renormalisation, after we include $N_G = (8\pi G_4)^{1/2}/2\pi$ (where $G_4$ is Newton's constant)  is given by
\begin{equation}
\delta \supset -  \frac{4g_s^2 \beta G_4}{3\pi}  \log |k^2\kt^2| .
\end{equation}
Note that when we sum over all equivalent diagrams and thus remove the field theory singularity, the $\log \kt^2$ term still remains. However,
this is of course not a singularity, as we have up to this point been
considering $\kt\gg \alpha'$. 

As $\kt$ decreases, the amplitude should smoothly revert to the correction for $\theta = 0$. To find the deviation from Newtonian behaviour at large distances we are interested in the variation of $\delta$ for small $\kt^2/\ap$, which as discussed above will be dominated by the same terms as in the large limit; for the term we have been considering we obtain
\begin{eqnarray}
L_{\delta}^{FT} &=& g_s^2 N_G^2 \frac{\beta \tilde{k}^2}{16\pi^2 \ap} \int_{\pi \ap}^{\infty} \frac{\d T}{T}  \int_0^1 \d y (1-2y)^2 e^{-4k^2 T y(1-y)}  + O((\frac{\tilde{k}^2}{4\pi^2 \ap})^2) \\
&=& \frac{g_s^2 \beta \tilde{k}^2 G_4}{8\pi^3 \ap} 2\int_0^1 \d y \ y(1-y)(1 - \gamma_E - \log4\pi - \log(\ap k^2 y(1-y))) + O((\ap k^2)^2) \nonumber 
\end{eqnarray}
from which we extract the contribution to the renormalisation:
\begin{equation}
\delta \supset - \frac{g_s^2 \beta G_4}{24 \pi^3 }\frac{\kt^2}{\ap}
\left(-\frac{5}{3} + \gamma_E + \log 4\pi +\log \ap k^2 \right).
\end{equation}

\section{Phenomenology: modification of gravity at a $mm$}

We now turn to phenomenological issues beginning briefly with the
possibility of Lorentz violation in the photon. In the Introduction
we mentioned birefringence of the trace U(1) photon which is constrained
by astrophysical observations. Taking into account our analysis and
the fact that the Lorentz violating operator $\Pi_{2}$ vanishes in
a fully supersymmetric theory, the velocity shift is of order \beq
\Delta v\sim c\frac{\lambda M_{SUSY}^{2}M_{s}^{2}}{M_{NC}^{4}}.\eeq
Following ref.\cite{Kostelecky:2001mb,Kostelecky:2002hh} a relatively
firm constraint comes from {}``time of flight'' signals from pulsars;
\beq
\frac{\sqrt{\lambda}\, M_{SUSY}M_{s}}{M_{NC}^{2}}\sim\sqrt{\lambda}\,\frac{M_{SUSY}}{\Lambda_{IR}}<2\times10^{-8},\eeq
where $\lambda$ is here a measure of the one-loop suppression in
the gauge diagrams, and $M_{SUSY}$ is a measure of the supersymmetry
breaking. A natural question to ask is how low the IR cut-off can
be; in other words, {\em is it likely that a regime that is well approximated
by noncommutative gauge theory will ever be accessible?} Alas, the answer
is no. Since $\lambda$ is a loop suppression factor involving
known gauge couplings it will be at least of order $10^{-3}$ assuming
that the mixing between the physical photon and trace U(1) photon
is of order unity. However supersymmetry is broken and transmitted,
one should almost certainly take $M_{SUSY}>1TeV$ giving \beq
\Lambda_{IR}>10^{9}GeV.\eeq
This bound is comparable to those coming from atomic physics calculated
in ref.\cite{Mocioiu:2001nz}; \beq
M_{NC}>10^{14}GeV.\eeq
Assuming that $M_{s}<M_{Pl}$, that bound translates into \beq
\Lambda_{IR}>2\times10^{10}GeV.\eeq
 If the physical photon has significant mixing with the trace U(1)
photon, it seems likely therefore that a non-zero $B$ field would
be felt as residual Lorentz violation rather than full blown noncommutative
field theory. For more detailed discussion of these questions 
see ref.\cite{Abel:2006sp}. 

Consider instead the possibility that the physical photon does not
mix with the trace U(1) photon. This could be the case if the trace
U(1) photon forms part of a hidden sector, or if the trace U(1) is
spontaneously broken by for example a Fayet-Iliopoulos term, if it
is anomalous. In this case $M_{NC}$ can be much lower and a significant
effect can show up in gravitional interactions. Our general analysis shows
that the graviton two point function in a theory with
nonvanishing $\theta$ tends continuously to the commutative one with
leading terms suppressed by factors of $\tilde{k}^{2}/\alpha'$. Neglecting
the possible implications of a non-trivial tensor structure for the
moment, the mildest effect one expects is a modification of the Newtonian
force law which derives from it. The observable effects will make
themselves felt as we probe the gravitational interaction at shorter
distances. As we saw, there is something akin to 
a {}``nonplanar'' one-loop contribution in the sense that 
$\tilde{G}(\mathbf{k})$ interpolates between
the $\tilde{k}^{2}\gg\alpha'$ regime and the $\tilde{k}^{2}\ll\alpha'$
regime where it deviates from the purely commutative model as $\tilde{k}^{2}/\alpha'$.
Neglecting tensor structure, we can therefore model the two point
function as \beq
\tilde{G}(\mathbf{k})=\frac{1}{M_{Pl}^{2}\mathbf{k^{2}}}\,\frac{1+f(\frac{\mathbf{\tilde{k}^{2}}}{\alpha'})}{1+\lambda}\eeq
where $f(x)\rightarrow\lambda(1+{\cal O}(x))$ for $x\ll1$ and tends
to the short range behaviour for $x\gg1$. 
Here $M_{Pl}^{2}$ is
the one loop Planck mass, which includes also tree level disk diagram contributions
such as those considered in ref. \cite{Alvarez-Gaume:2006bn}.
For example if we assume that the one-loop
contribution has power law behaviour $\sim|\tilde{k}|^{(3-p)}$ we
can model the total tree and one-loop two point function as \beq
\tilde{G}(\mathbf{k})=\frac{1}{M_{Pl}^{2}\mathbf{k^{2}}}\frac{1}{(1+\lambda)}\left(1+\lambda\left(\frac{1}{1+\frac{\mathbf{\tilde{k}^{2}}}{\alpha'}}\right)^{\frac{p-3}{2}}\right).\eeq
 The coefficient $\lambda$ encapsulates the one-loop open string
contribution to Planck's constant in the commutative theory with $\theta=0$,
which can be significant and is model dependent. Indeed there are
generic scenaria that lead to the extremes $\lambda\gg1$ and $\lambda\ll1$:

\begin{enumerate}
\item The ADD scenario \cite{Antoniadis:1997zg,Antoniadis:1998ig}: 
the Standard Model is associated with a local brane
configuration (for example in a {}``bottom-up'' construction as
per the previous section), with the 4D Einstein-Hilbert action deriving
from the dimensionally reduced 10D action. In this case the one loop
correction will be localized whereas the large tree-level $M_{Pl}^{2}$
is the result of a large volume. The one loop open string contribution
will therefore be suppressed by a factor \beq
\lambda\sim\frac{1}{V_{10-p}}\eeq
where $V_{10-p}$ is the extra-dimensional volume in units of $\sqrt{\alpha'}$.
3-branes in the original ADD scenario with TeV scale gravity would
therefore lead to a tiny $\lambda$, but one could imagine the Standard
Model localized on wrapped D7-branes for example, in which case intermediate
values of $\lambda$ are possible.

\item The DGP scenario \cite{Dvali:2000hr}: 
gravity is localized to a 3-brane in infinite or
large extra dimensions by one-loop diagrams with matter (brane) states
in the loop. The novel feature is that gravity becomes higher dimensional
at long distances, offering an explanation of the observed cosmological
acceleration. In this case one expects $\lambda\gg1$
in the region where gravity is 4 dimensional. In more detail, the
full action consists of a bulk term and a one loop induced brane term;
\beq
M_{Pl}^{2}\left(\int d^{4}x\sqrt{g_{4}}R^{(4)}+\rho_{c}^{D-4}\int d^{D}x\sqrt{g_{D}}R^{(D)}\right),\eeq
where $R^{(4)}$ is the curvature form the induced metric on the brane.
Since $\rho_{c}^{D-4}$ appears in the propagator with a factor $k^{2}$
it is natural that the cross-over length scale \emph{above} which
gravity appears $D$ dimensional, generically given by \cite{Kiritsis:2001bc}
\beq
R_{c}=\alpha'^{\frac{6-D}{4}}\rho_{c}^{\frac{D-4}{2}}.\eeq
This possibility has been analyzed for (Type I) open string models
in ref.\cite{Antoniadis:1996vw,Epple:2004ra,Kiritsis:2001bc}, where in practice a number
of different threshold effects are possible if the matter branes wrap
some compact internal dimensions. The precise details of these other
thresholds will not change our conclusions about the effect of UV/IR
mixing.
\end{enumerate}
To see the effect of the one-loop corrections on the potential between
two point particles consider for example $\theta^{12}=\theta$. In
this case \beq
\mathbf{\tilde{k}}^{2}  =  \theta^{2}(\mathbf{k}_{1}^{2}+\mathbf{k}_{2}^{2})
\,  = \,  \theta^{2}\mathbf{k}^{2}\sin^{2}\vartheta \, , \eeq
where $\vartheta$ is the angle to the 3 direction. The potential
depends on the angle $\vartheta$ and is given by the retarded Green's
function;\begin{eqnarray}
V(\mathbf{x}) & = & \int dt\, G_{R}(t,\mathbf{x}) \nn \\
 & = & \int\frac{d^{3}\mathbf{k}}{(2\pi)^{3}}\tilde{G}(\mathbf{k})e^{i\mathbf{k}.\mathbf{x}}\end{eqnarray}
which leads to \beq
V(r,\vartheta)=\frac{1}{8\pi M_{Pl}^{2}r}\,\left(1+\frac{1}{1+\lambda}\int_{0}^{\infty}\left(f\left(\frac{r_{c}^{2}}{r^{2}}\, y^{2}\right)-\lambda\right)\, e^{-y\cos\vartheta}J_{0}(y\sin\vartheta)\, dy\right)\eeq
where \beq
r_{c}=\frac{\theta}{\sqrt{\alpha'}}=\frac{M_{s}}{M_{NC}^{2}}.\eeq
In the limit where $r\cos\vartheta\gg r_{c}$ we may expand $f$ inside
the integral. Using the identity \beq
\int_{0}^{\infty}y^{m\geq0}e^{-y\cos\vartheta}J_{0}(y\sin\vartheta)\, dy=(-1)^{m}m!\, P_{m}(\cos\vartheta)\eeq
we find that the leading deviation from Newtonian behaviour is a quadrupole
moment that sets in at $r\sim r_{c}$: indeed if $f(x)=\lambda(1+\beta x+\ldots)$
we find \beq
V(r,\vartheta)=\frac{1}{8\pi M_{Pl}^{2}r}\left(1+\frac{\lambda\beta\,(3\cos^{2}\vartheta-1)}{(1+\lambda)}\,\frac{r_{c}^{2}}{r^{2}}+\mathcal{O}\left(\frac{r_{c}^{4}}{r^{4}}\right)\right).\eeq
The radius $r_{c}$ is the distance above which Planck's constant
tends to the $B=0$ one-loop value. This is a potential which can
be compared directly with the experimental bounds presented in ref.\cite{expt}.
Also note that there is a direction given by $\cos\vartheta=0$ where
the physics is identical to $\theta=0$ physics. 

At smaller distances the {}``nonplanar'' contribution to Planck's
constant diminishes. For $r\ll r_{c}$ we may use the identity
\beq
e^{-y\cos\vartheta}J_{0}(y\sin\vartheta)\,=\sum_{n=0}^{\infty}\frac{(-1)^{n}y^{n}}{n!}P_{n}(\cos\vartheta)\eeq
and approximate\begin{eqnarray}
f\left(\frac{r_{c}^{2}}{r^{2}}\, y^{2}\right) & = & \lambda\left(\frac{1}{1+\frac{r_{c}^{2}}{r^{2}}\, y^{2}}\right)^{\frac{p-3}{2}}\end{eqnarray}
to find the first few harmonics as \beq
V(r,\vartheta)=\frac{1}{8\pi M_{Pl}^{2}r}\left(\frac{1}{1+\lambda}+\sum_{n=0}^{p-4}\frac{(-1)^{n}B(\frac{p-n-4)}{2},\frac{n+1}{2})}{2n!}P_{n}(\cos\vartheta)\left(\frac{r}{r_{c}}\right)^{1+n}+\mathcal{O}\left(\frac{r}{r_{c}}\right)^{p-2}\right).\eeq
The leading term is the tree-level Planck's constant, and the subleading
terms grow with radius, as they should, to build up the full one-loop
Planck's constant at large distance. 

The most notable general conclusion from this analysis is simply that
the distance scale at which the modification of gravity takes place,
\beq
r_{c}=\frac{\theta}{\sqrt{\alpha'}}=\frac{M_{s}}{M_{NC}^{2}},\eeq
can be much larger than the inherit distance scales in the model.
For example if $M_{s}\sim M_{Pl}$ and $M_{NC}\sim1TeV$ then $r_{c}\sim1mm$
(the same numerical coincidence as the large extra dimension scenarios
with 2 extra dimensions).

\section{Conclusions}

Noncommutative field theory provides a theoretical framework to discuss
effects of nonlocality and Lorentz symmetry violation. Proper
understanding and better control of the 
UV/IR mixing has been a serious obstacle for the field theory. 
In this paper, we have emphasised that the IR singularities are 
just a reflection of the fact that field theory is UV divergent. 
Consequently any attempt to resolve them
without modifying the UV
behaviour of the field theory is doomed, and they 
can only be consistently smoothed out in a UV finite theory.
We have demonstrated
this explicitly by considering noncommutative field theory as an  
approximation to open string theory with a background $B$-field. We showed 
that the noncommutative field theory description is valid only for the
intermediate range of energy scale $ \Lambda_{IR}^2 \equiv \alpha'
M_{NC}^2 < k^2< 1/\alpha'$ and explored what happens outside this range. 
The IR singularities are rendered harmless and in
fact, long before they are reached, the singular 
IR physics of the noncommutative 
theory is replaced by regular physics that is dictated by the UV 
finiteness of strings. 
In many non-supersymmetric theories, tachyonic instabilities arise 
from the modified dispersion relation (\ref{Dispersion}) \cite{Ruiz:2000hu}, 
which our analysis implies are also resolved by embedding into an
UV-complete theory, as discussed in the context of field theory in \cite{Abel:2006sp}.

With the UV/IR mixing under control, one can now reliably study how
noncommutative geometry modifies the IR physics. Below the
noncommutative IR scale $\Lambda_{IR}$, normal Wilsonian behaviour is
resumed and the low energy physics can be described in terms of
ordinary local physics with residual Lorentz violating operators.
Indeed the theory tends continuously to the commutative $B=0$
field theory, with the Lorentz violating operators 
remaining as a footprint in the low energy phenomenology 
of the string scale physics.
A second important example of how the low energy physics 
is modified arises in the gravitational sector. We studied how the 
noncommutative geometry may modify gravity by
considering the graviton two point function. 
The departure from the ordinary Newtonian potential can be much more 
significant and happen at much lower energy scales than those suggested 
by any extra dimensions. 

One aspect of the present study that requires further elaboration is the 
nature of the effective field theory in the gravity sector and 
the resulting cosmology. Because of the difficulty of extracting an 
effective field theory for the gravitational sector it is 
not clear how these features will turn out, or indeed if they 
lead to any strong observational constraints.

\section*{Acknowledgements}   

We thank Joerg Jaeckel, Valya Khoze, Hong Liu, Andreas Ringwald, Rodolfo Russo and Gary Shiu for very useful discussions and comments.
The research of CSC is supported by EPSRC through an Advanced
Fellowship. 

\appendix

\section{Field Theory Limits of String Diagrams - a Review}

First we divide the Schwinger integrals as described above so that
\beq
\Pi_1(k,-k)=\Pi_{1_{IR}}+\Pi_{1_{UV}},\eeq
where 
UV and IR indicate $t\in[0,1]$ and $t\in[1,\infty]$
respectively. Considering the $IR$ contribution of the planar diagram
note that, if we reduce $\theta\rightarrow0$, then the field-theory limit 
should be the same for planar and non-planar diagrams; equivalently, 
they should be the same up to $O(\ap)$ corrections. This is not 
immediately obvious from the Green's functions, but we must bear 
in mind that the planar diagrams have spurious poles on the worldsheet, 
and the string amplitude is strictly only defined after analytically 
continuing the momentum \cite{DiVecchia:1996uq,DiVecchia:1996kf}. 
The field theory limit is obtained by taking $t \gg 1$ and excising 
the regions around the poles - i.e. the region $|x - x^{\prime}| < 1$ and 
$t - |x-x^{\prime}| < 1$ - and then keeping terms of lowest order in $w$:
\begin{equation}
I^{P,NP}_0 = -\frac{(x-x^{\prime})^2}{t} + \pi |x-x^{\prime}| \pm 
\Delta + O(e^{-2\pi t}),
\end{equation} 
where $\Delta$ is of order $w$, and the $-(+)$ preceding it applies 
to planar (non-planar) diagrams. We retain this term due to the presence 
of the tachyon, as in \cite{DiVecchia:1996uq,Bilal:2000bk}; it is given by
\begin{equation}
\Delta = e^{-2\pi x} + e^{2\pi (x-t)}
\end{equation}
so that $\dot{\Delta}^2 = -4\pi^2 + O(w)$, but for the superstring we shall 
find that it is irrelevant. Inserting the above into (\ref{Pi1B}) and 
extracting the contribution from the first level in the loop, we find
\begin{multline}
\Pi_{1_{IR}}^{P}(k^2)  =  -4\alpha^{\prime 2} g_{D_p}^{2}\int_{1}^{\infty}\d t \,(8\pi^{2}\alpha't)^{-\frac{(p+1)}{2}}e^{2\pi t}(1+(24-d)\, e^{-2\pi t}+\ldots)\times \\
\int_0^{t} \d x e^{-2\ap k^2 \pi(x - \frac{x^2}{t})} \left[ \frac{-2\pi x}{t} + \pi + \dot{\Delta} + ... \right]^2 \\
= \frac{- g_{D_p}^{2}}{(4\pi)^{\frac{(p+1)}{2}}} \int_{2\pi \ap}^{\infty}\d T \ T^{-\frac{(p-1)}{2}}  \int_0^{1} \d y \ e^{- T k^2 (y - y^2)} \left[ (24-d) (1 - 2y)^2 - 8 + ... \right].
\end{multline}
This result looks just like the field theoretical Schwinger integral
as it should (note the change to the parameters $T= 2\pi \ap t$ and $y = x/t$). 
We have not explicitly written the tachyonic contribution
or contributions coming from states at higher excitation level: the
tachyon because it is unphysical, and the higher states because their
nonplanar counterparts in the IR ($p\rightarrow0$) are all finite.
For the moment we need only note that a contribution at level $n$
yields a Schwinger integral of the form \beq
\int_{2\pi \alpha'}^{\infty}dT\, T^{-\frac{(p-1)}{2}}\int_{0}^{1}dy\,(1-2y)^{2}e^{-T(k^{2}(y-y^{2})+(n-1)\alpha'^{-1})}.\eeq
 
To obtain the field theory limit, we perform the integrals above and then take the $\ap \rightarrow 0$ limit; we can do this using the exponential integral. For example, when $p=3$ we have the standard field theory behaviour, with
\begin{equation}
\Pi_{1_{IR}}^P = \frac{d}{3} \ \frac{g_{D_3}^{2}}{(4\pi)^{2}} \ln k^2 + O(1).
\label{Pi1PD}\end{equation}
For $d=22$ we obtain the beta function of ref.\cite{Bilal:2000bk}, but for the case $d=0$, we find that the leading logarithm cancels, and we have the finite result
\begin{equation}
\Pi_{1_{IR}}^P = \frac{16}{3} \ \frac{g_{D_3}^{2}}{(4\pi)^{2}} + O(k^2).
\label{Pi1PD0}\end{equation}

\end{document}